\documentclass{sig-alternate-hgk}
\usepackage{url}
\usepackage{mahnigCEC}
\usepackage{epsfig}
\begin{document}
%
  \conferenceinfo{ } { }
    \CopyrightYear{2014}
    \crdata{}
    \clubpenalty=10000
    \widowpenalty = 10000

\title{A probabilistic evolutionary optimization approach to compute quasiparticle braids}

%
%
%
%
%

\numberofauthors{4} 
%
\author{
%
%
\alignauthor
Roberto Santana
 \affaddr{Intelligent Systems Group}\\
 \affaddr{Dept. of Computer Science and Artificial Intelligence}\\
 \affaddr{University of the Basque Country (UPV/EHU)}\\
 \email{roberto.santana@ehu.es}
\alignauthor
 Ross B. McDonald\\
 \affaddr{Dept. of Physics and Astronomy}\\
 \affaddr{Texas A\&M University, College Station, TX 77843-4242, USA}\\
       \email{}
\alignauthor 
 Helmut G. Katzgraber
 \affaddr{Dept. of Physics and Astronomy}\\
 \affaddr{Texas A\&M University, College Station, TX 77843-4242, USA}\\
 \affaddr{Santa Fe Institute, 1399 Hyde Park Road, Santa Fe, NM 87501, USA}\\ 
       \email{katzgraber@physics.tamu.edu}
}
\date{30 July 1999}

\maketitle
\begin{abstract}
  Topological quantum computing is an alternative framework for avoiding the quantum decoherence problem in quantum computation. The problem of executing a gate in this framework can be posed as the problem of braiding quasiparticles. Because these are not Abelian, the problem can be reduced to finding an optimal product of braid generators where the optimality is defined in terms of the gate approximation and the braid's length. In this paper we propose the use of different variants of estimation of distribution algorithms to deal with the problem. Furthermore, we investigate how the regularities of the braid optimization problem can be translated into statistical regularities by means of the Boltzmann distribution. We show that our best algorithm is able to produce many solutions that approximates the target gate with an accuracy in the order of $10^{-6}$,  and  have lengths up to $9$ times shorter than those expected from braids of the same accuracy obtained with other methods.   
\end{abstract}


\section{Introduction}

   The idea of using the theory of quantum mechanics to obtain computers potentially exponentially faster for certain applications, such as the factorization of prime numbers, arouses considerable interest and research efforts from the scientific community nowadays. In quantum computation, information is represented and manipulated using quantum properties. An obstacle for the construction of large quantum computers is the problem of quantum decoherence, that can be viewed as the loss of information of the quantum system due to the interaction with the environment. It terms of the computation goal, quantum decoherence can be also understood as unwanted noise introduced in quantum computation \cite{Sarma_et_al:2006}. One possible solution to  this problem is the design of quantum systems immune to quantum decoherence on a hardware level. 

 Topological quantum computing (TQC) \cite{Bonesteel_et_al:2005,Sarma_et_al:2006} investigates quantum computing systems that, given the properties of quasiparticles they use, are not affected by quantum decoherence.  The key idea of these systems is that quantum information can be stored in global properties of the system and thus affected only by global operations but not by local perturbations such as noise. In TQC, quantum gates are carried out by adiabatically braiding quasiparticles around each other. This braiding is used to perform the unitary transformations of a quantum computation. 

 One of the essential questions to design a TQC is to find a product of braid generators (matrices) that approximates a quantum gate with the smallest possible error and, if possible, as short as possible to prevent loss \cite{Mcdonald_and_Katzgraber:2013}.  Some approaches that treat this question as an optimization problem have been proposed. Exhaustive search  \cite{Bonesteel_et_al:2005} has been applied to search for braids of manageable size (up to $46$ exchanges). More recently, McDonald and Katzgraber \cite{Mcdonald_and_Katzgraber:2013} have proposed the use of genetic algorithms to find optimal braids. They also introduce a function that takes into account the goals of maximizing the accuracy and minimizing the length. In this paper, we build on their results to propose an analysis of the braid optimization problem using probabilistic modeling of the space of braid solutions. For braid problems of small size, we show that the regularities that exist in the search space can be captured by the probabilistic models. We then extend these results to propose the application of evolutionary algorithms (EAs) able to capture statistical regularities of the best solutions.
 
 Estimation of distribution algorithms (EDAs) \cite{Larranhaga_and_Lozano:2002r,Muhlenbein_and_Paas:1996r,Larranaga_et_al:2012} are EAs that apply learning and sampling of distributions instead of classical crossover and mutation operators. Modeling the dependencies between the variables of the problem can serve to efficiently orient the search to more promising areas of the search space by explicitly capturing and exploiting potential relationships between the problem variables. In addition, the information contained in the models can reveal previously unknown information about the problem.  

 The paper is organized as follows:  The next section briefly describes braids and Fibonacci anyons  \cite{Bonesteel_et_al:2005}.  The problem formulation, including the problem representation and fitness functions used are presented in Section~\ref{sec:PROBFORM}. The framework for probabilistic analysis of braids is introduced in Section~\ref{PROBBRAIDS}. This section also includes a number of experiments that help to illustrate the rationale of our approach. Section \ref{sec:EDAs} presents the different variants of EDAs proposed for the braids problem.  Work related to our approach is discussed in Section~\ref{sec:RELWORK}. Section~\ref{sec:EXPE} describes the experimental framework to evaluate our proposal and  presents the numerical results of our experiments. The main contributions of the paper are summarized in Section~\ref{sec:CONCLU} where some lines for future research are also discussed.

\section{Braids and anyons} \label{sec:BRAIDS}

Qubits play in quantum computation a role similar to that played by bits in digital computers. A braid operation can be represented by a matrix that acts on the qubit space. These matrices are referred to as generators, and the quantum gate that a braid represents is the product of the generators that encode the individual braid operations. 

Let $\sigma_1$ and $\sigma_2$ represent two possible generators.  $\sigma_1^{-1}$   and $\sigma_2^{-1}$ respectively represent their inverses. Given a braid $B$, $len()$ is a function that  returns the braid's length $l$ (e.g. $B=\sigma_1 \sigma_1 \sigma_2 \sigma_1^{-1}$, $l=len(B)=4$).

 Since the product of a matrix by its inverse  reduces to the identity matrix, some braids can be simplified reducing their length. Therefore, we also define function $elen()$, that has a braid as its argument and returns the braid's \emph{effective length} which is the length of braid after all possible simplifications have been conducted.

\begin{align}
  \sigma_1 \sigma_1 \sigma_1 \sigma_1 \sigma_1^{-1} =&   \sigma_1 \sigma_1 \sigma_1 \label{eq:EL1} \\
  \sigma_2^{-1} \sigma_1 \sigma_1 \sigma_1^{-1} \sigma_1^{-1}  \sigma_2 \sigma_1^{-1} =& \sigma_1^{-1} \label{eq:EL2} 
\end{align}

 In the braids shown in examples \eqref{eq:EL1} and  \eqref{eq:EL2}, the effective length values are $3$ and $1$, respectively.

Let $T$ represent the target matrix (gate to be emulated), the braid error is calculated with the following metric  \cite{Mcdonald_and_Katzgraber:2013}:
\begin{equation}
 \epsilon = |B-T| \label{eq:ERROR}
\end{equation}
where the matrix norm used is
\begin{equation}
 |M| = \sqrt{\sum_{ij}M_{ij}^2}.
\end{equation}

The problem of finding braiding operations that approximate gates is then reduced to finding a product chain of the reduced generators and their inverses that approximates the matrix representing the quantum gate. Two elements that describe the quality of a braid are its error $\epsilon$ and its length $l$. 

\subsection{Fibonacci anyon braids}

 Anyons appear as emergent quasiparticles in fractional quantum Hall states and as excitations in microscopic models of frustrated quantum magnets that harbor topological quantum liquids \cite{Read_and_Rezayi:1999}. Fibonacci anyons are the simplest anyons with non-Abelian braiding statistics that can give rise to universal quantum computation. Fibonacci anyon braids \cite{Bonesteel_et_al:2005} only encompasses one-qubit gates. In such systems, the braid transition operators result in a phase change for the non computational state, and therefore it can be ignored. Overall, phases in the problem can also be ignored. Therefore the transition matrices can be projected onto SU(2) by a multiplication with $e^{\frac{i \pi}{10}}$, yielding for the generators

\begin{equation}
\sigma_1 =  \left( \begin{array}{cc}
e^{\frac{-i7\pi}{10}}   & 0 \\
0 & -e^{\frac{-i3\pi}{10}}   \end{array} \right) \label{eq:SIG1}
\end{equation}

\begin{equation}
\sigma_2 =  \left( \begin{array}{cc}
-\tau e^{\frac{-i \pi}{10}}   & -i \sqrt{\tau} \\
-i \sqrt{\tau} & -\tau e^{\frac{-i\pi}{10}}   
\end{array} \right) \label{eq:SIG2}
\end{equation}
where $\tau = \frac{\sqrt{5} -1}{2}$. 

 In this paper we address the problem of finding a product of generator matrices for Fibonacci anyon braids. Although the methodology we propose can be extended to other braids, we focus on anyon braids since they are one of the best known in TQC \cite{Mcdonald_and_Katzgraber:2013,Xu_and_Xin:2008}. As a target gate for computing the error \eqref{eq:ERROR} we use 

\begin{equation}
T= \left( \begin{array}{cc}
i   & 0 \\
0 & i   
\end{array} \right).
\end{equation}

\section{Problem formulation} \label{sec:PROBFORM}

\subsection{Problem representation}

Let ${\bf{X}}=(X_1,\ldots ,X_n)$ denote a vector of discrete random variables. We use ${\bf{x}}=(x_1,\ldots ,x_n)$ to denote an assignment to the variables. $I$ denotes a set of indices in $\{1, \ldots, n\}$, and $X_I$ (respectively $x_I$) a subset of the variables of ${\bf{X}}$ (respectively ${\bf{x}}$) determined by the indices in $I$.

 In our representation for the quasiparticle braids problem,  ${\bf{X}}=(X_1,\ldots ,X_n)$ represents a braid of length $n$, where  
$X_i$ takes values in $\{0,1, \dots, 2g-1\}$ and $g$ is the number of generators. Given an order for the generators $\sigma_1, \sigma_2, \dots, \sigma_g$,  $X_i=j, j<g$ means that the matrix in position $i$ is $\sigma_{j+1}$. If $X_i =j, j \geq g$, then the matrix in position $i$ is $\sigma^{-1}_{(j-g)+1}$.  For example, for generators shown in Equations~\eqref{eq:SIG1} and~\eqref{eq:SIG2}, and  $B=\sigma_1 \sigma_1 \sigma_2 \sigma_2^{-1} \sigma_1^{-1}$, the corresponding braid representation is  ${\bf{x}}=(0,0,1,3,2)$.

\subsection{Fitness function}

We are interested in the solution of an optimization problem formulated as   ${\bf{x}}^* = arg max_{{\bf{x}}} f({\bf{x}})$,  where $f : S \rightarrow R$ is called the objective or fitness function. The optimum   ${\bf{x}}^*$ is not necessarily unique.

To evaluate the fitness function associated to a solution ${\bf{x}}$, firstly the product of braid matrices $B$ is computed according to ${\bf{x}}$ and then the error $\epsilon$ is calculated from $B$ as in~\eqref{eq:ERROR}. 

The fitness function  \cite{Mcdonald_and_Katzgraber:2013} is defined as:
\begin{equation}
 f({\bf{x}}) = \frac{1-\lambda}{1+\epsilon} + \frac{\lambda}{l} \label{eq:FITNESS}
\end{equation}
where $l$ is the braid's length, and the parameter $\lambda$ serves to balance the two conflicting goals, i.e., having short braids or low approximation error. When $\lambda=0$, braids are optimized only for the error and the function reaches its maximum value when this error is minimized. 

 We define functions $\hat{f}({\bf{x}})$ and  $\bar{f}({\bf{x}})$ as two variations of function~\eqref{eq:FITNESS}. Function  $\hat{f}({\bf{x}})$ is identical to $f({\bf{x}})$, except that the effective length $\hat{l}=elen(B)$ is used instead of the braid's length. Function $\bar{f}({\bf{x}})$ outputs the maximum value of the function for any of the braids contained in $B$ that start from the first position, i.e. 
\begin{equation}
 \bar{f}({\bf{x}}) =  max_{{\bf{y}}, {\bf{y}} \in \{(x_1),(x_1,x_2),(x_1, \dots,x_i), (x_1,\dots,x_n)\}} f({\bf{y}}) 
\end{equation}

\section{Probabilistic modeling of braids} ~\label{PROBBRAIDS}

 To optimize the braid problem we  use a class of evolutionary algorithms that capture and exploit statistical regularities in the best solutions. These methods assume that such regularities exist.  As a preliminary proof of concept on the existence of such regularities, we investigate the Boltzmann distribution for braids of manageable size.  A similar approach has been successfully applied to investigate the dependencies that arise in the configurations of simplified protein models \cite{Santana_et_al:2008a} and conductance-based neuron models \cite{Santana_et_al:2012h}.

\subsection{Boltzmann distribution}

 When the dimension of the braid problem is small, complete enumeration and evaluation of all possible solutions is feasible. In this situation, brute force can be applied to identify the optimal solutions. We use complete enumeration to define a probability distribution on the space of all possible braids for $n = 10$. Using the fitness value as an energy function, we associate to each possible braid  a probability value  $p({\bf{x}})$ according to the Boltzmann probability distribution.  The Boltzmann probability distribution  $p_B({\bf{x}})$ is defined as

  \begin{equation}
  p_B({\bf{x}}) = \frac{e^{\frac{g({\bf{x}})}{T}}}{
  \sum_{{\bf{x}}'} e^{\frac{g({\bf{x}}')}{T}}}, \label{eq:BOLTPROB}
  \end{equation}
  where $g({\bf{x}})$ is a given objective function and $T$ is the system temperature that can be used as a parameter to smooth the the probabilities.

The Boltzmann probability distribution  is used in statistical physics to associate a probability with a system state according to its energy \cite{VanKampen:1992}.  In our context of application, $p_B({\bf{x}})$ assigns a higher probability to braids that gives a more accurate approximation to the target gate. The solutions with the highest probability correspond to the braids that maximize the objective function. We use an arbitrary choice of the  temperature, $T=1$, since our idea is to compare the distributions associated to different fitness functions for the same parameter $T$.

\begin{figure*}[hbp]
  \begin{center}
    \includegraphics[width=5.5cm]{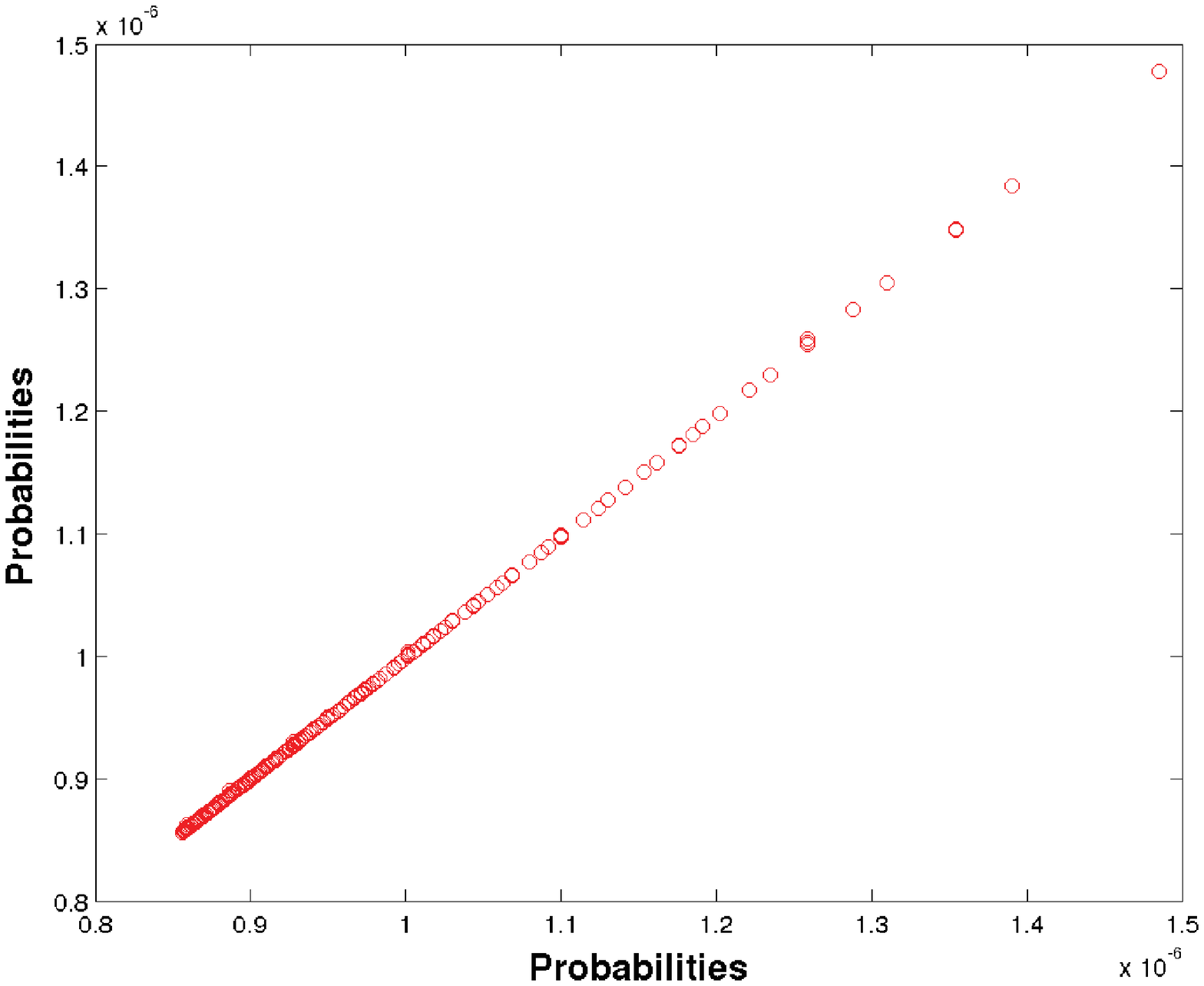}
    \includegraphics[width=5.5cm]{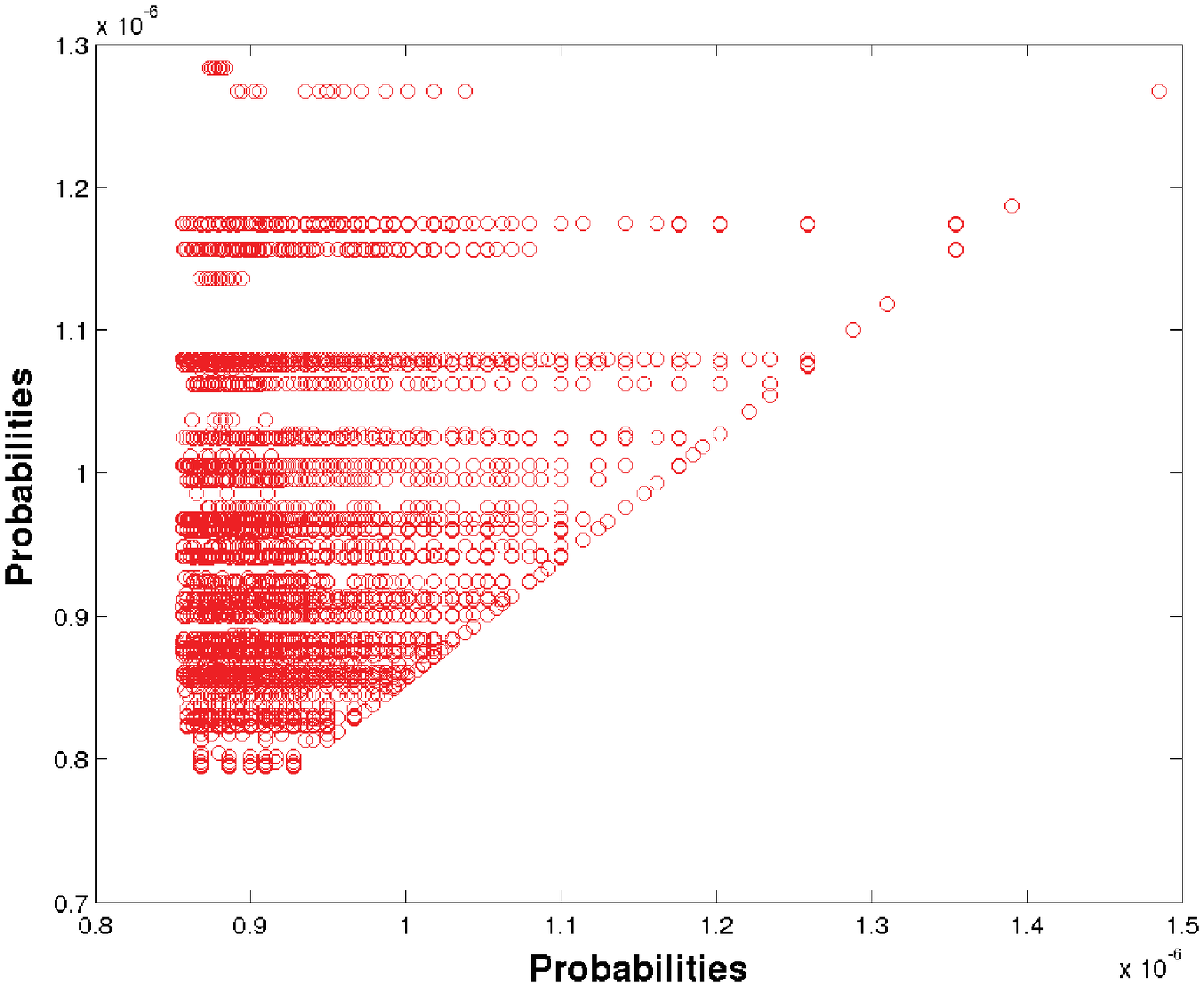}
    \includegraphics[width=5.5cm]{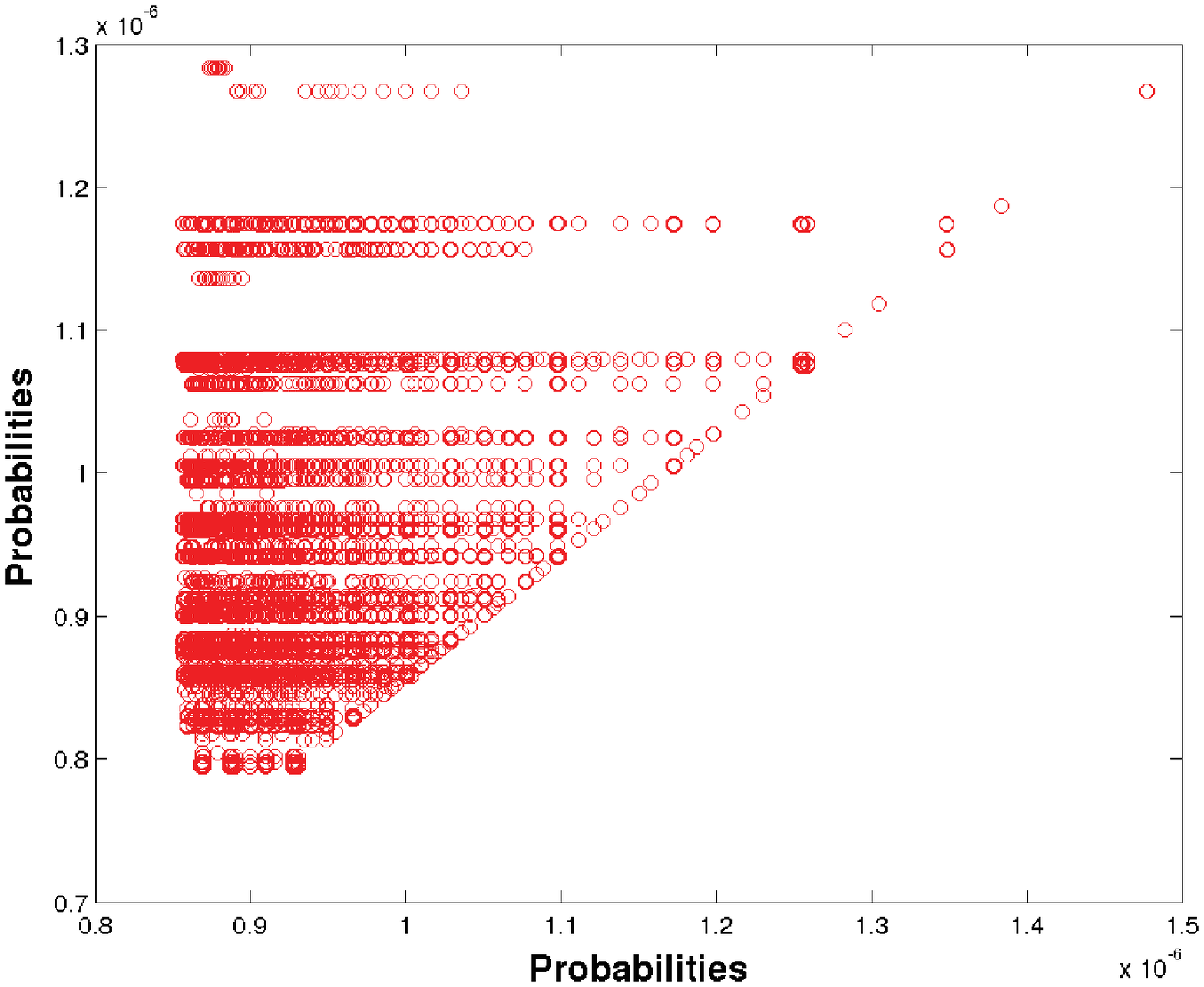}
\begin{pspicture}(0,-0.25)(12,0)
    \rput(1,-0.125){a)} \rput(6.2,-0.125){b)} \rput(12.0,-0.125){c)}
\end{pspicture}
     \caption{Boltzmann distribution computed for different functions. a) $f$ vs  $\hat{f}$ , b) $f$ vs $\bar{f}$, and c)  $\hat{f}$ vs $\bar{f}$.}
    \label{fig:BOLT}
    \end{center}
\end{figure*}

 Using the Boltzmann distribution we can investigate how potential  regularities of the fitness function are translated into statistical properties of the distribution. In particular, we investigate the marginal probabilities associated to the variables and the mutual information between pairs of variables.  Figure~\ref{fig:BOLT} shows the probabilities assigned by  the Boltzmann distribution to functions $f$, $\hat{f}$, $\bar{f}$, for $n=10$. The search space comprises  $4^{10}=1,048,576$ braids. 

 It can be seen in Figure~\ref{fig:BOLT}a) that probabilities assigned by the Boltzmann distribution to function  $f$ and  $\hat{f}$ are very similar although not identical. For both functions, only few points have a high probability. The important difference between functions  $f$ and  $\bar{f}$ is evident in Figure~\ref{fig:BOLT}b). The probability assigned by the Boltzmann distribution to a braid is always higher or equal for function $\bar{f}$ than for function $f$. The reason is that $\bar{f}$ considers a greater space of solutions. Differences between functions  $f$ and  $\hat{f}$ can be also detected by comparing figures~\ref{fig:BOLT}b) and~\ref{fig:BOLT}c). For the three functions, the space of solutions with lower fitness is more dense that the solutions with higher fitness.

\subsection{Statistical analysis of the braids space}

\begin{figure*}[hbp]
  \begin{center}
     \includegraphics[width=5.5cm]{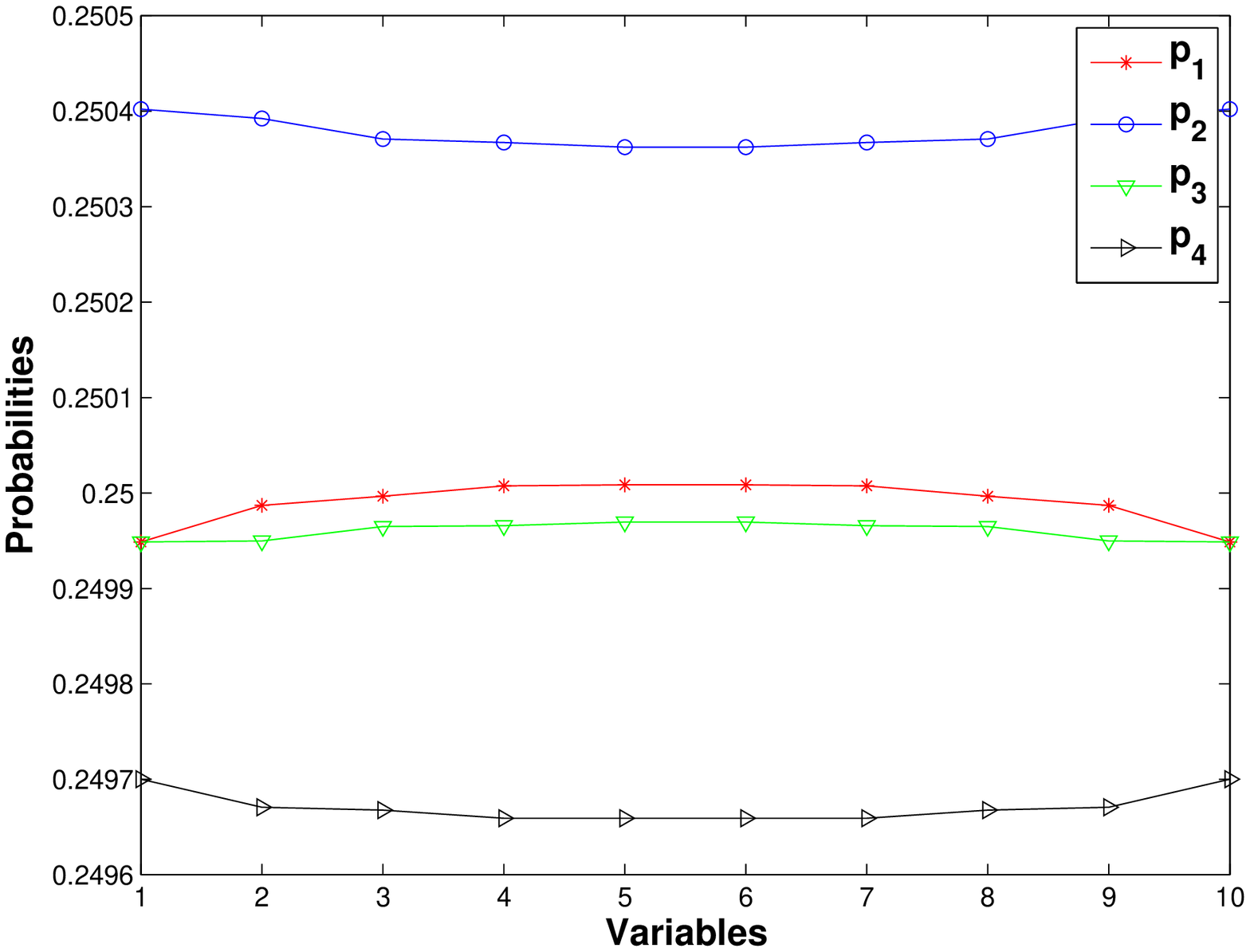}
     \includegraphics[width=5.5cm]{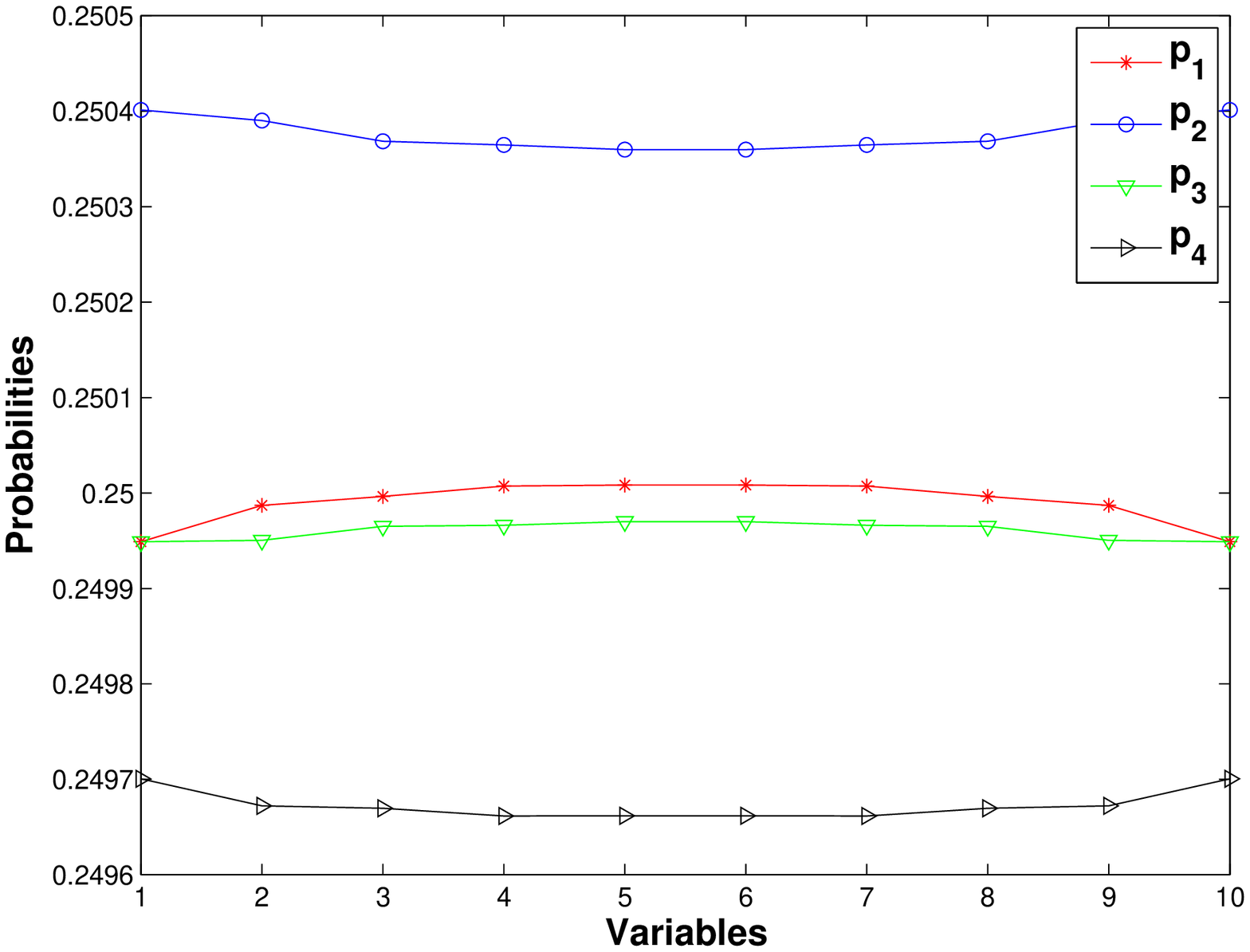}
     \includegraphics[width=5.5cm]{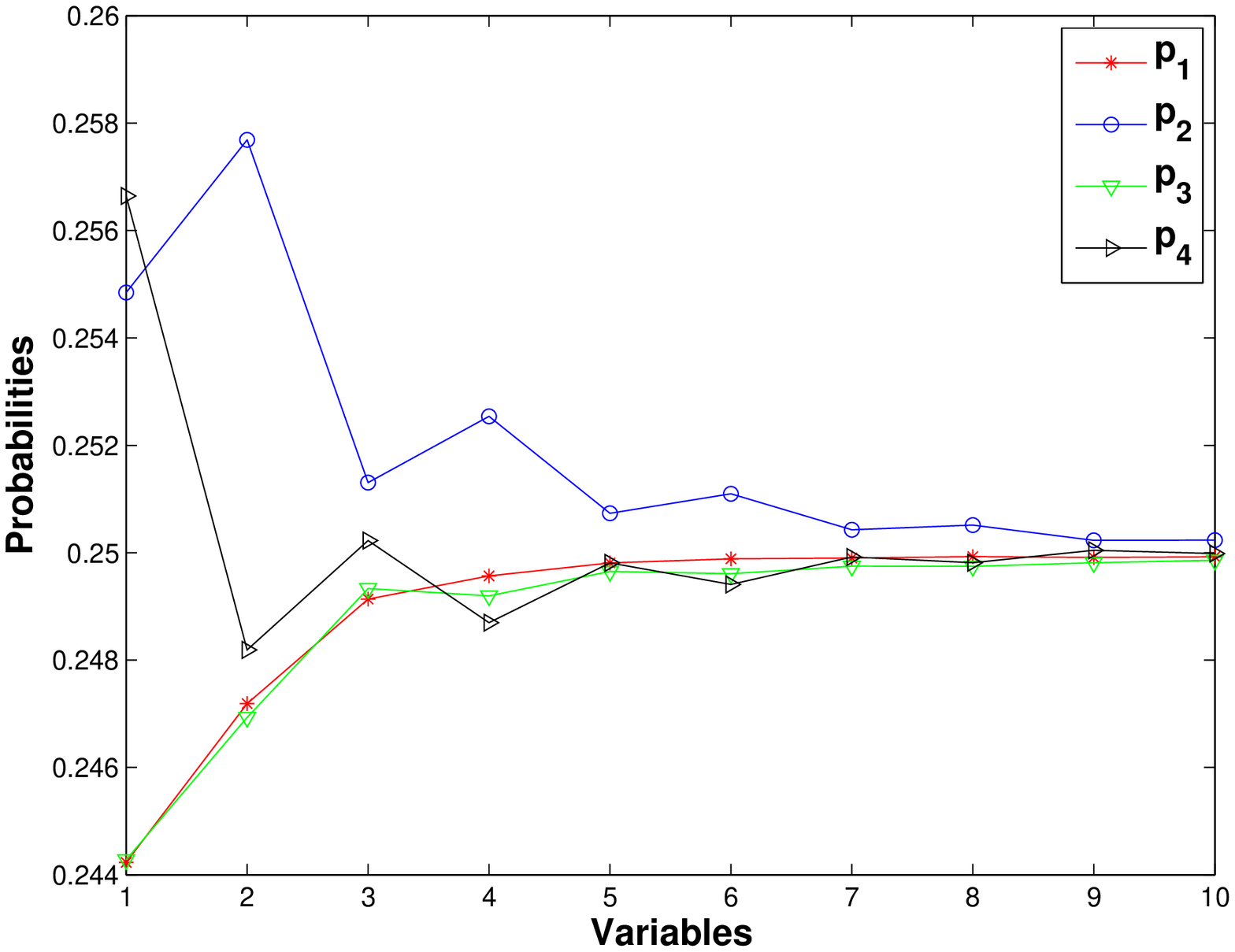}
\begin{pspicture}(0,-0.25)(12,0)
    \rput(1,-0.125){a)} \rput(6.2,-0.125){b)} \rput(12.0,-0.125){c)}
\end{pspicture}
     \caption{Univariate probabilities computed from the Boltzmann distribution for :  a) $f$ , b) $\hat{f}$, and c) $\bar{f}$.}
     \label{fig:UNIV}
    \end{center}
\end{figure*}

 Figure~\ref{fig:UNIV} shows the univariate probabilities computed from the Boltzmann distribution for the three functions analyzed and the $10$ variables. $p_1$, $p_2$, $p_3$, and $p_4$ respectively represent the univariate probabilities for braid generators $\lambda_1$, $\lambda_2$, $\lambda_1^{-1}$, and $\lambda_2^{-1}$. For all the functions, higher probabilities for $p_3$ indicate that $\lambda_1^{-1}$ is more likely to be present in the best solutions. This is the type of statistical regularities that can be detected and exploited by EAs that learn probabilistic models. If a particular configuration is more likely to be present in the best solutions, then, these configurations could be sampled more frequently at the time of generating new solutions.

  Analysis of Figure~\ref{fig:UNIV} also reveals the similarities between functions $f$ and  $\hat{f}$ since they determine similar univariate distributions for all the variables. A remarkable fact is that for function  $\bar{f}$ (Figure~\ref{fig:UNIV}c)) the univariate probabilities  notably differ for the first variables and are more similar as the index of the variables increases. One possible explanation for this behavior is that the first variables are present in more solutions of those considered by function  $\bar{f}$. Changes in these variables are more influential in the function. Finally, another remarkable observation is that the univariate probabilities associated to $\lambda_1$ and $\lambda_1^{-1}$ are always very close for the three functions. 

\begin{figure*}[hbp]
  \begin{center}
    \includegraphics[width=5.5cm]{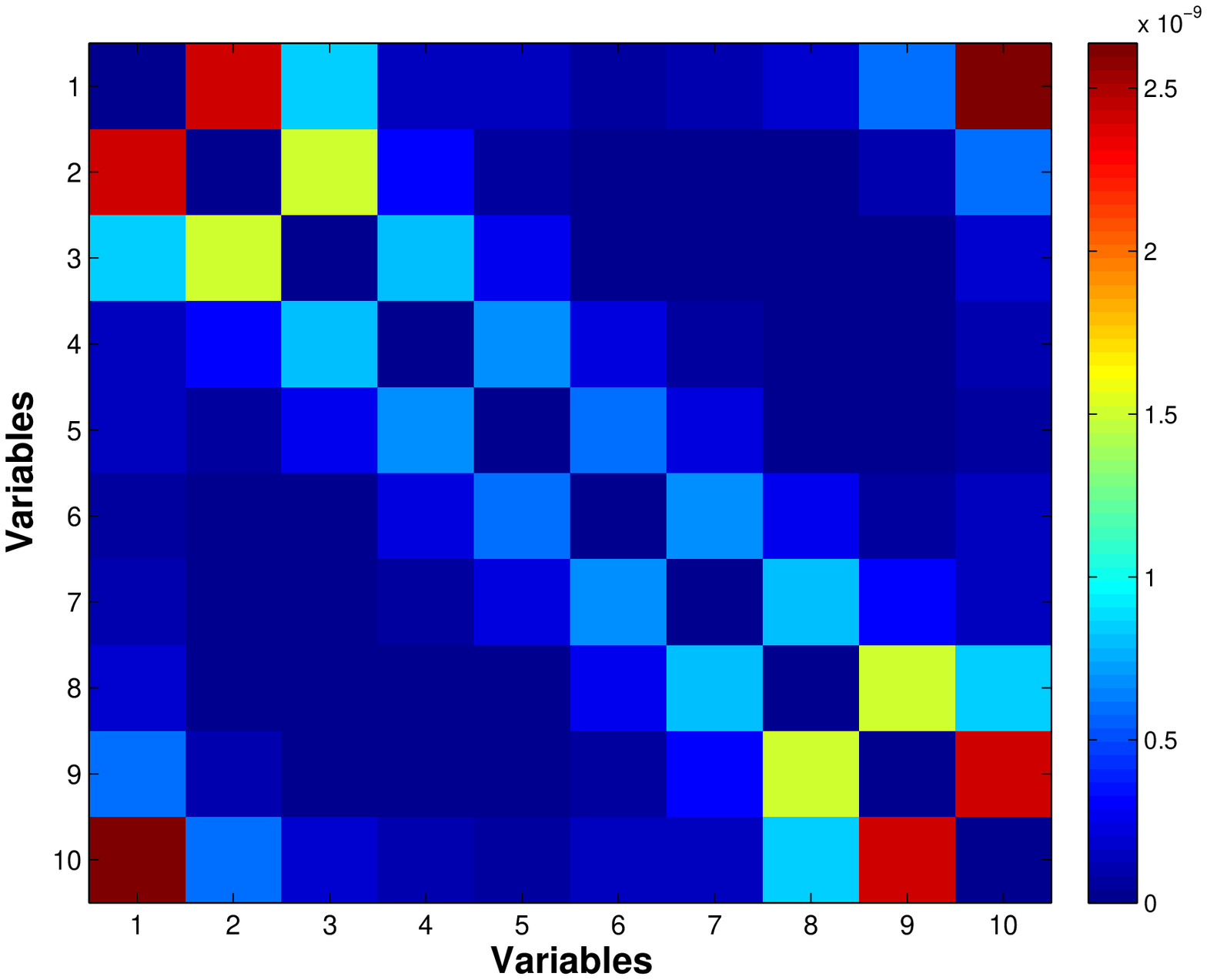}
    \includegraphics[width=5.5cm]{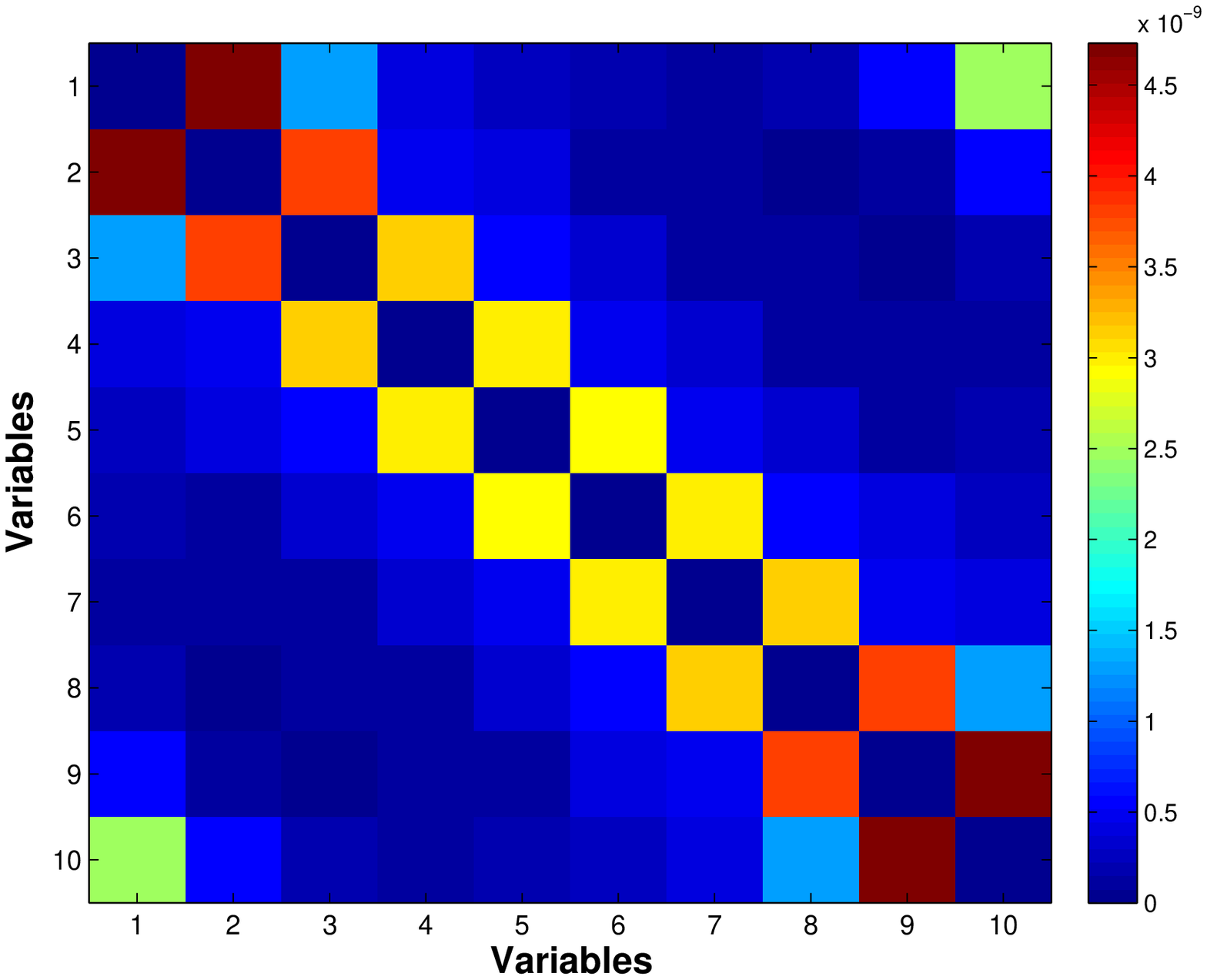}
    \includegraphics[width=5.5cm]{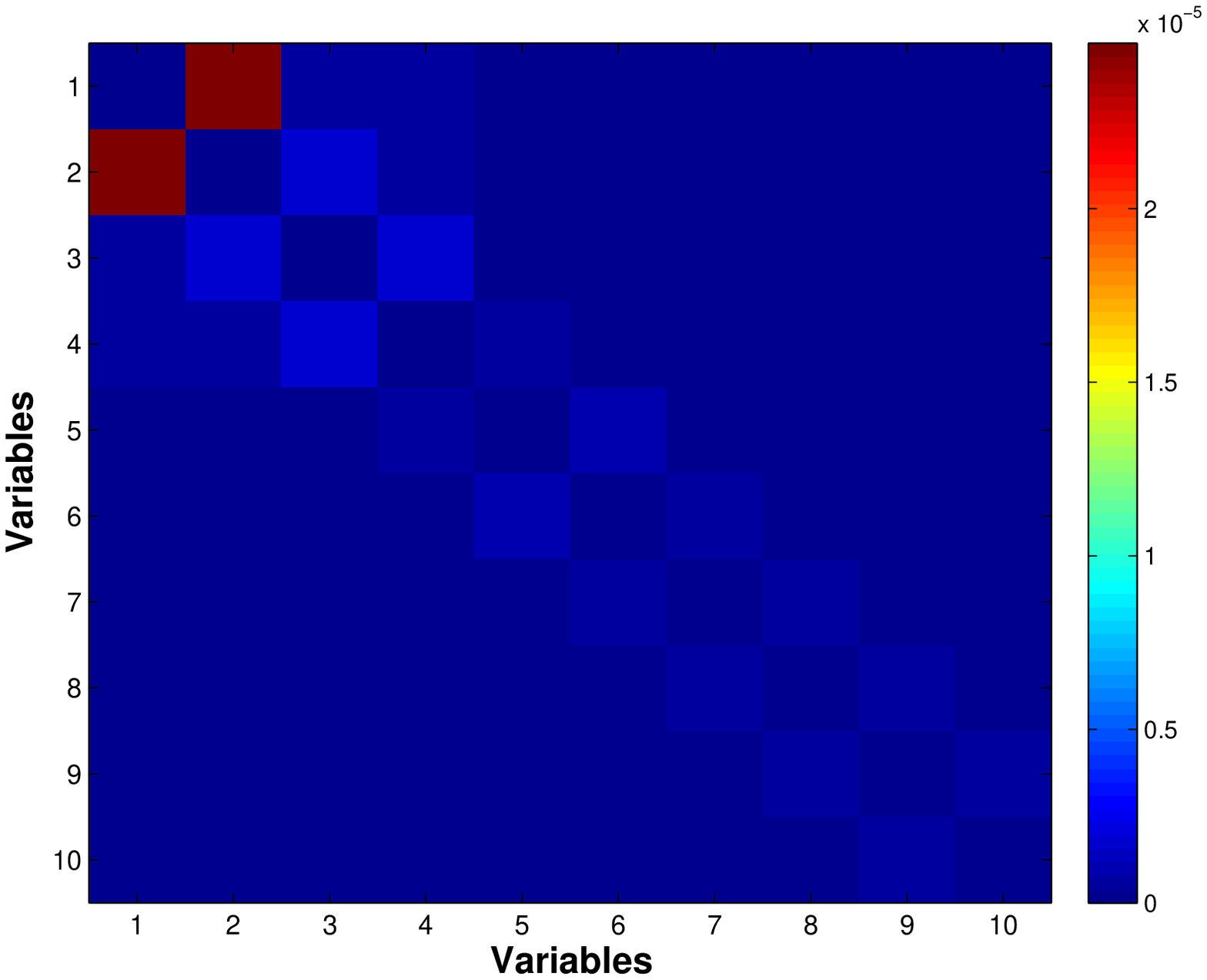}
\begin{pspicture}(0,-0.25)(12,0)
    \rput(1,-0.125){a)} \rput(6,-0.125){b)} \rput(12.0,-0.125){c)}
\end{pspicture}
     \caption{Mutual information computed from the Boltzmann distribution for: a) $f$, b) $\hat{f}$, and c) $\bar{f}$.}
    \label{fig:MI}
    \end{center}
\end{figure*}

Finally, using the Boltzmann probabilities, we compute the bivariate marginal distributions between every pair of variables and derive the values of the mutual information. The mutual information is a measure of statistical dependence between the variables and can serve to identify variables that are dependent. A strong dependence between two variables may indicate that their joint effect has a strong influence on the function. Figure~\ref{fig:MI} shows the mutual information computed for the three functions analyzed. 

It can be seen in Figure~\ref{fig:MI} that for the three functions the strongest dependencies are between adjacent variables, although for functions $f$ and  $\hat{f}$ there is also a strong dependence between the first variable and the last variable. Although the pattern of dependence is similar in functions  $f$ and  $\hat{f}$, the mutual information is higher for function  $\hat{f}$. It can also appreciated in Figure~\ref{fig:MI}c) that the dependencies between adjacent variables decreases with the index for function $\bar{f}$. 

 Summarizing, the statistical analysis of the Boltzmann distribution shows that there are at least two types of regularities of the braid problem that are translated into statistical features. Firstly, there are different frequencies associated to the generators in the space of the best solutions. Secondly, there are strong dependencies between the variables, particularly those that are adjacent in the braid representation.

\begin{table*} 
\begin{center}
 {$ \begin{array}{|c|l|} \hline 
n &  \multicolumn{1}{c|}{\text{braid}}  \\ \hline  \hline
50 &  \sigma_1^{-1}\sigma_2^{3}\sigma_1^{-2}\sigma_2\sigma_1^{-1}\sigma_2\sigma_1^{-2}\sigma_2^{3}\sigma_1^{-1}\sigma_2\sigma_1^{-1}\sigma_2\sigma_1^{-2}\sigma_2\sigma_1^{-2}\sigma_2\sigma_1^{-1}\sigma_2^{3}\sigma_1^{-1}\sigma_2^{2}\sigma_1^{-1}\sigma_2\sigma_1^{-1}\sigma_2\sigma_1^{-2}\sigma_2^{2}\sigma_1^{-3}\sigma_2^{2}  \\  \hline

100 & \sigma_1\sigma_2^{-2}\sigma_1^{4}\sigma_2^{-1}\sigma_1\sigma_2^{-4}\sigma_1\sigma_2^{-1}\sigma_1^{3}\sigma_2^{-1}\sigma_1^{2}\sigma_2^{-2}\sigma_1^{4}\sigma_2^{-1}\sigma_1\sigma_2^{-4}\sigma_1\sigma_2^{-1}\sigma_1^{6}\sigma_2^{2}\sigma_1\sigma_2\sigma_1\sigma_2^{2}\sigma_1^{3}\sigma_2^{5}\sigma_1^{-1}\sigma_2\sigma_1^{-3}\sigma_2\sigma_1^{3}\sigma_2^{5} \\ \hline

150 & \sigma_2\sigma_1^{-1}\sigma_2\sigma_1^{-1}\sigma_2^{-1}\sigma_1^{2}\sigma_2^{-1}\sigma_1\sigma_2^{-4}\sigma_1^{2}\sigma_2^{-2}\sigma_1^{-1}\sigma_2^{-1}\sigma_1^{-2}\sigma_2^{-4}\sigma_1^{2}\sigma_2^{-4}\sigma_1\sigma_2^{-1}\sigma_1^{2}\sigma_2^{-1}\sigma_1^{-1}\sigma_2\sigma_1^{-3}\sigma_2^{-1}\sigma_1\sigma_2^{-1}\sigma_1^{3}\sigma_2^{-1}\sigma_1\sigma_2^{-4}\sigma_1^{2}\sigma_2^{-4}\sigma_1^{2}\sigma_2^{-3}\\ \hline

200 & \sigma_2^{-2}\sigma_1^{4}\sigma_2^{-1}\sigma_1\sigma_2^{-4}\sigma_1\sigma_2^{-1}\sigma_1^{2}\sigma_2^{-1}\sigma_1^{3}\sigma_2^{-1}\sigma_1\sigma_2^{-4}\sigma_1\sigma_2^{-1}\sigma_1^{4}\sigma_2^{-1}\sigma_1\sigma_2^{-2}\sigma_1^{2}\sigma_2^{2}\sigma_1\sigma_2^{-1}\sigma_1\sigma_2^{-1}\sigma_1^{-5}\sigma_2\sigma_1^{-1}\sigma_2^{-6}\sigma_1^{-2}\sigma_2^{3} \\  \hline

 250 & \sigma_1\sigma_2\sigma_1^{-1}\sigma_2\sigma_1^{-1}\sigma_2^{-1}\sigma_1\sigma_2^{-1}\sigma_1^{-3}\sigma_2^{4}\sigma_1^{-2}\sigma_2^{4}\sigma_1^{-1}\sigma_2\sigma_1^{2}\sigma_2\sigma_1^{-1}\sigma_2^{4}\sigma_1^{-2}\sigma_2^{4}\sigma_1^{-3}\sigma_2^{-1}\sigma_1\sigma_2^{-1}\sigma_1^{-1}\sigma_2^{2}\sigma_1\sigma_2^{-1}\sigma_1^{2}\sigma_2^{-1}\sigma_1\sigma_2^{-4} \sigma_1^{2}\sigma_2^{-4}\sigma_1\sigma_2^{-1}\sigma_1^{4}\\
& \sigma_2^{-1}\sigma_1^{-1}\sigma_2^{-1}\sigma_1^{-4}\sigma_2^{-1}\sigma_1^{-1}\sigma_2^{-1}\sigma_1^{4}\sigma_2^{-1}\sigma_1\sigma_2^{-4}\sigma_1^{2}\sigma_2^{-4}\sigma_1\sigma_2^{-2}\sigma_1^{-1}\sigma_2^{5}\sigma_1^{-1}\sigma_2\sigma_1^{-4}\sigma_2^{2}\sigma_1^{-4}\sigma_2^{2}\sigma_1^{-4}\sigma_2^{2}\sigma_1^{-1} \\ \hline
\end{array}$}
\end{center}
\caption{Best braids found by the EDAs for each value of $n$.}
\label{tab:BESTSOL}
\end{table*}

\section{Modeling the braid space}  \label{sec:EDAs}

Using  the Boltzmann distribution to find the statistical regularities is not feasible for real problems for which the space of solutions can not be inspected exhaustively. However, statistical regularities can be detected in samples of solutions selected according to their fitness. EDAs use samples of solutions to learn a model that captures some of the regularities that may exist in the data. The pseudocode of an EDA is shown in Algorithm~\ref{alg:EDA}.
 
\begin{BAlgo}{Estimation of distribution algorithm}
 \label{alg:EDA}
 \item Set $t\Leftarrow 0$. Generate $N$ solutions randomly.
 \item \Do
 \item \T {Evaluate the solutions using the fitness function.}
 \item \T {Select a population $D_t^S$ of $K \leq N$ solutions according to a selection
method.}
 \item \T {Calculate a probabilistic model of $D_t^S$.}
 \item \T {Generate $N$ new solutions sampling from the distribution represented
in the model.}
 \item \T {$t \Leftarrow t+1$}
 \item  \Until{Termination criteria are met.}
\end{BAlgo}

  The model used by the EDA will determine the type of regularities that will be captured from the data and also the cost of the optimization algorithm since more complex models are generally more computationally costly.

 
  In this paper we use three types of probabilistic graphical models: 1) Univariate model. 2) $1$-order Markov model. 3) Tree model. We work  with positive distributions denoted by $p$. $p(x_I)$ denotes  the  marginal probability for ${\bf{X}}_I={\bf{x}}_I$.   $p(x_i \mid x_j)$ denotes the conditional probability distribution of $X_i=x_i$ given $X_j=x_j$.  

In the univariate model variables are considered to be independent, and the probability of a solution is the product of the univariate probabilities for all variables:

\begin{equation}
 p_{u}({\bf{x}}) =   \prod_{i=1}^{n}  p(x_{i}) \label{eq:UNIV}
\end{equation}

In the $1$-order Markov model, the configuration of variable $X_i$ depends on the configuration of its previous variable:

\begin{equation}
 p_{MK}({\bf{x}}) =  p(x_{1})  \prod_{i=2}^{n}  p(x_{i} \mid  {x_{i-1}}) \label{eq:MKFACT}
\end{equation}

  A probability distribution $p_{\mathcal{T}}({\bf x})$ that is conformal with a tree is defined as:
\begin{equation}
  p_{\mathcal{T}} ({\bf{x}}) =\prod_{i=1}^{n} p(x_i|pa(x_i)),
\end{equation}
 where $Pa(X_i)$  is the parent of $X_i$ in the tree, and $p(x_i|pa(x_i))=p(x_i)$ when $pa(X_i)=\emptyset$, i.e. $X_i$ is a root of the tree. We allow the existence of more than one root in the PGM (i.e. forests) although for convenience of notation we refer to the model as tree. 
 
 Univariate approximations are expected to work well for functions that can be additively decomposed into  functions of order one (e.g. $g({\bf{x}})= \sum_i x_i$). However, other non additively decomposable functions can be easily solved with EDAs that use univariate models (e.g. $g({\bf{x}})=   \prod_i x_i + \sum_i x_i $) \cite{Muhlenbein_et_al:1999}. Therefore, it makes sense to test the univariate approximation for the braid problem.  The 1-order Markov model  captures only dependencies between adjacent variables, and the tree model can represent a maximum of $n-1$ bivariate dependencies. 

 EDAs that used univariate, one-order Markov, and tree models  were respectively introduced in \cite{Baluja:1994,Muhlenbein_and_Voosen:1993a}, \cite{DeBonet_et_al:1996r} and~\cite{Baluja_and_Davies:1997r} and details on the methods used to learn and sample the models can be obtained from these references. 

 \subsection{Enhancements to the EDAs} \label{sec:ENH}

   We consider three enhancements to EDAs: 1) Use of a local optimizer. 2) Partial sampling. 3) Recoding. 
 
   As is the case of other EAs, EDAs can be enhanced by the incorporation of local optimizers \cite{Pelikan_and_Goldberg:2003r}. We use a greedy optimization algorithm that is applied during the evaluation of the population by the EDA.  The algorithm starts from the solution generated by the EDA. In each iteration, the local optimizer evaluates all the $3n$ solutions that are different to the current solution in only one variable (the neighbor solutions). The next selected solution is the neighbor that improves  the fitness of the current solution the most. The algorithm stops when none of the neighbors improves the fitness of the current solution. 
 
  During the sampling step of an EDA, all variables are assigned their values according to the probabilistic model and the sampling method. For the EDA that uses the univariate model, variables are independently sampled. For 1-order Markov and tree, probabilistic logic sampling (PLS) \cite{Henrion:1988} is used. In both methods, all variables are assigned the new values. However, for some problems with higher-order interactions using a base template solution can be better than generating each new solution from scratch.  For the braid problem, careful recombination of the solutions proved to be essential for the genetic algorithm introduced in \cite{Mcdonald_and_Katzgraber:2013}.

 In partial sampling, a solution of the population is selected and only a subset of its variables are sampled according to the model. We use two variants of partial sampling I) Partial sampling where the number of variables to be modified is randomly selected between $1$ and $n$. II)  Partial sampling, where  the number of variables to be modified is randomly selected between $1$ and $\frac{n}{2}$.

  Recoding consists in modifying the representation of the solution according to the evaluation. For functions $\hat{f}$ and $\bar{f}$ it is possible to recode the solution by eliminating redundant generators (e.g.,  pairs $\sigma_i \sigma_i^{-1}$). The rationale of using recoding is that meaningful variables will be located closer to the beginning of the braid. Since solutions have a fixed length, the last variables will be kept unused, i.e. garbage information. Therefore, we devised two ways to fill these gaps: I) Leaving the unused variables as they were in the original solution. II) Replacing the unused variables by a reverse copy of the variables used in the evaluation. The second variant intends to replicate information that has proved to be ``informative'' about the problem. Equations~\eqref{eq:EX3} and~\eqref{eq:EX4} show examples of recoding type I and II, respectively. In this hypothetical examples, the underlined variables are those that provided the best fitness after simplifying the braid and evaluating function $\bar{f}$.  

\begin{align}
   (\underline{0,3},1,3,\underline{3,3,2},1,2,2) =&  (\underline{0,3,3,3,2},3,2,1,2,2) \label{eq:EX3} \\
(\underline{0,3},1,3,\underline{3,3,2},1,2,2) =&  (\underline{0,3,3,3,2},2,3,3,3,0) \label{eq:EX4} 
\end{align}

\section{Related work} \label{sec:RELWORK}

 In addition to the use of genetic algorithms  \cite{Mcdonald_and_Katzgraber:2013}, brute force \cite{Bonesteel_et_al:2005} has been proposed to search for a braids of manageable size (up to $46$ exchanges).  Other methods such as the Solovay-Kitaev algorithm \cite{Dawson_et_al:2005,Hormozi_et_al:2007,Nielsen_and_Chuang:2010} provide bounds on the accuracy and length of the braids. However, these methods do not allow the user to  tune the balance between the accuracy and the length as pioneered in \cite{Mcdonald_and_Katzgraber:2013}.   

 The Boltzmann distribution has played an important role in the theoretical analysis of EDAs and other authors works have analyzed the relationship between the function structure and the dependencies in the distribution \cite{Muehlenbein_and_Mahnig:2000,Ochoa_and_Soto:2005,Santana:2005}. Other problems from physics have been previously treated with EDAs. In particular, spin glass models with different types of interactions and topologies have been addressed \cite{Pelikan_and_Goldberg:2003r,Pelikan_and_Katzgraber:2009,Shakya_et_al:2011}. Two important differences between the braid problem and the spin glass models that makes it particularly challenging is that its fitness function is multiplicative and that the representation is non binary. In fact the  cardinality of the variables can increase with the number of generators.

\section{Experiments} \label{sec:EXPE}

  The main objective of our experiments is to evaluate the capacity of the EDAs to find optimal solutions to the braid problem. We run experiments for $n \in \{50,100,150,200,250\}$ in order to evaluate the scalability of the algorithms.  A second objective is to compare different variants of the problem formulation and of the algorithm. 

\subsection{Experimental settings}

Each EDA is characterized by $5$ parameters:

\begin{itemize}
\item   Use of local optimizer. 0: Only EDA is applied, 1: EDA is combined with greedy search as described in Section \ref{sec:ENH}. 
\item  Type of function and representation. 0: Function $f$,  1: Function $\bar{f}$ without recoding, 2: Function $\bar{f}$ with recoding type I, 3: Function $\bar{f}$ with recoding type II.
 \item  $\lambda$ value.  0:0.0, 1:0.01, 2:0.05, 3:0.1. 
 \item    Sampling method. 0: Normal, 1: Partial sampling type I,  2: Partial sampling type II. 
 \item  Type of probabilistic model. 0: Univariate, 1: 1-order Markov, 2: Tree.
\end{itemize}

The total number of variants of the algorithm was $2 \times 4 \times 4 \times 3 \times 3 = 288$. All the algorithms use truncation selection, in which the best $5\%$ of the population is selected. EDAs that do not incorporate the greedy local search use a population size $N=10000$. For these EDAs, the number of generations  was dependent on $n$ as $N_g = 15n$. 

 Due to the large number of evaluations spent by the greedy search method, the population size for all hybrid EDAs was $N=100n$ and the number of generations was fixed to  $N_g = 100$. For each EDA variant, $100$ experiments were run.

\subsection{Best solutions found by EDAs}

\begin{table} 
\begin{center}
 {$ \begin{array}{|c|r|c|c|c|} \hline 
n & l & \epsilon & log_{10}(\epsilon) &  log_{10}^{3.97}(\frac{1}{\epsilon})  \\ \hline
50 & 44 & 4.8435 \times 10^{-4} & -3.3148 & 116.47\\  \hline
100 &70  &8.3527 \times 10^{-6} & -5.0782 & 633.37 \\ \hline
150 &64 & 8.3527 \times 10^{-6} & -5.0782 & 633.37 \\ \hline
200 &62 & 8.3527 \times 10^{-6} & -5.0782&  633.37 \\  \hline
250 &124 & 3.5038 \times 10^{-6} & -5.4555 & 841.82\\ \hline
\end{array}$}
\end{center}
\caption{Parameters of the best braids found by the EDAs for each value of $n$.}
\label{tab:BESTSOLPARAM}
\end{table}

Tables~\ref{tab:BESTSOL} and~\ref{tab:BESTSOLPARAM} respectively show the best braids achieved by the EDAs for each value of $n$ and the characteristics of these solutions. In Table~\ref{tab:BESTSOLPARAM}, we show an estimate of the length  of the braids ($O[log_{10}^{3.97}(1/\epsilon)]$) that would compute the Solovay-Kitaev algorithm to obtain the same error $\epsilon$ of our best solutions. The lengths of our solutions compare favorably with these estimates. 

\begin{figure}[htb]
  \begin{center}
 \includegraphics[width=6.0cm]{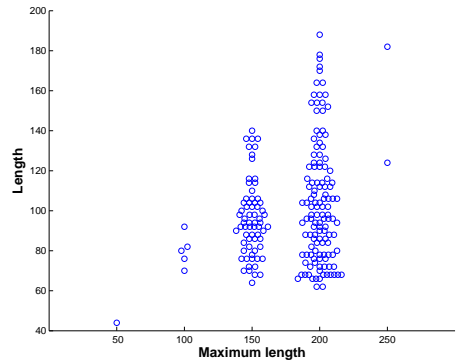}
  \caption{Length of the best solutions found for each value of $n$.}
    \label{fig:RHCSTEPS}   
 \end{center}
\end{figure}

  Figure~\ref{fig:RHCSTEPS} shows the length of all the best solutions found for each value of $n$. It can be observed in Figure~\ref{fig:RHCSTEPS}  that EDAs are able to find several braids with different lengths for $n=150$ and $n=200$. 

\subsection{Behavior of the different EDA variants}
 
\begin{figure*}[t]
  \begin{center}
    \includegraphics[width=5.3cm]{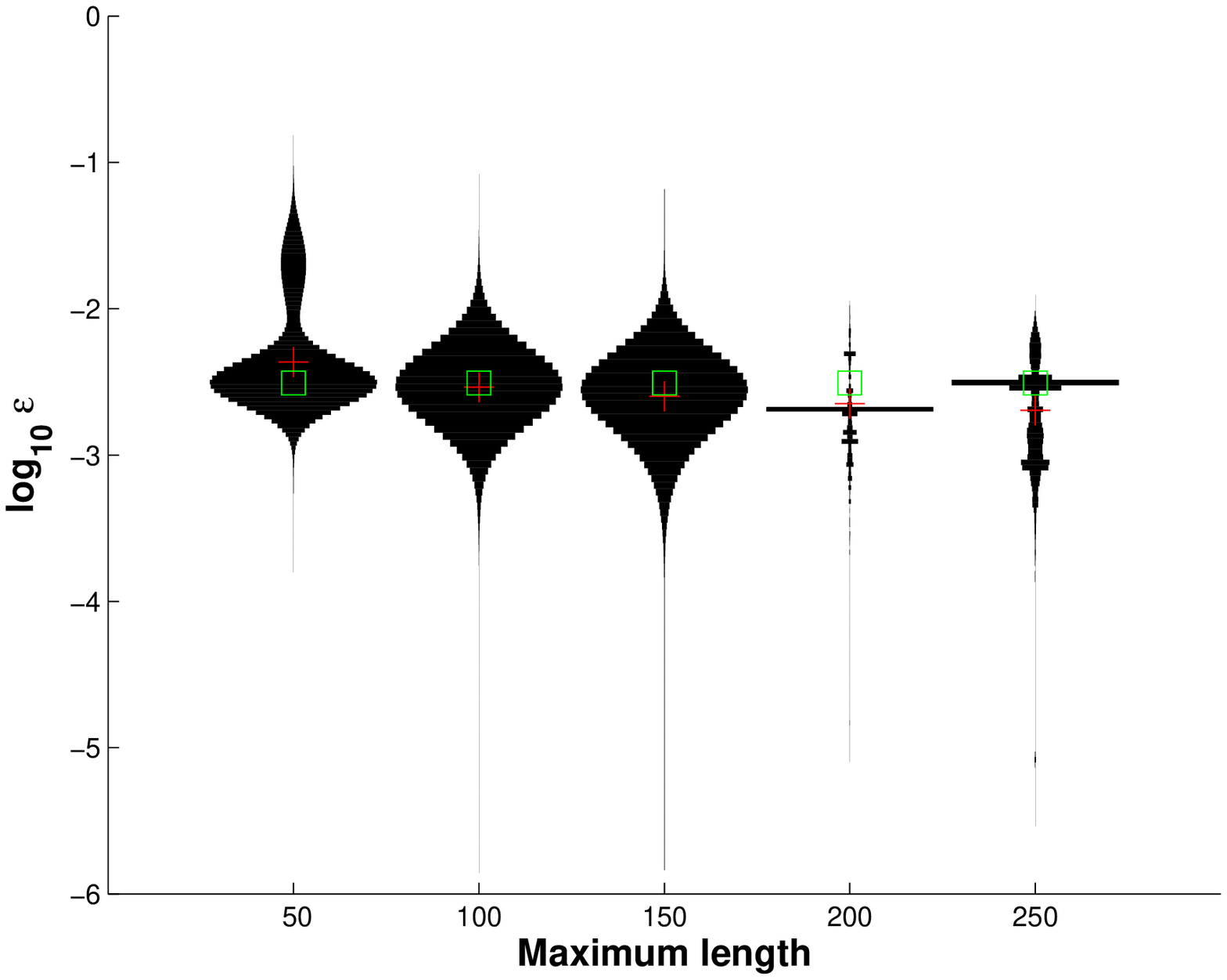}
    \includegraphics[width=5.3cm]{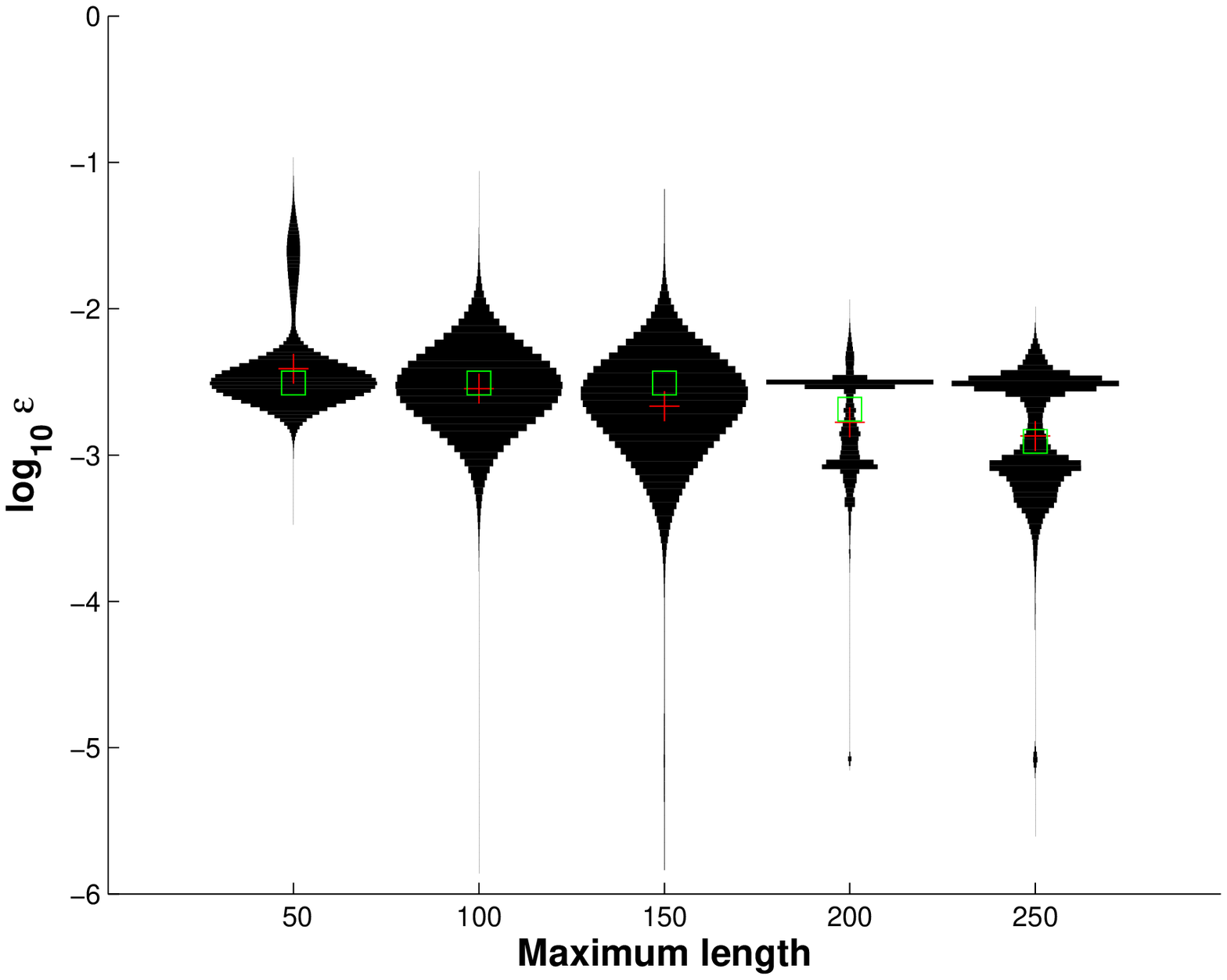}
    \includegraphics[width=5.3cm]{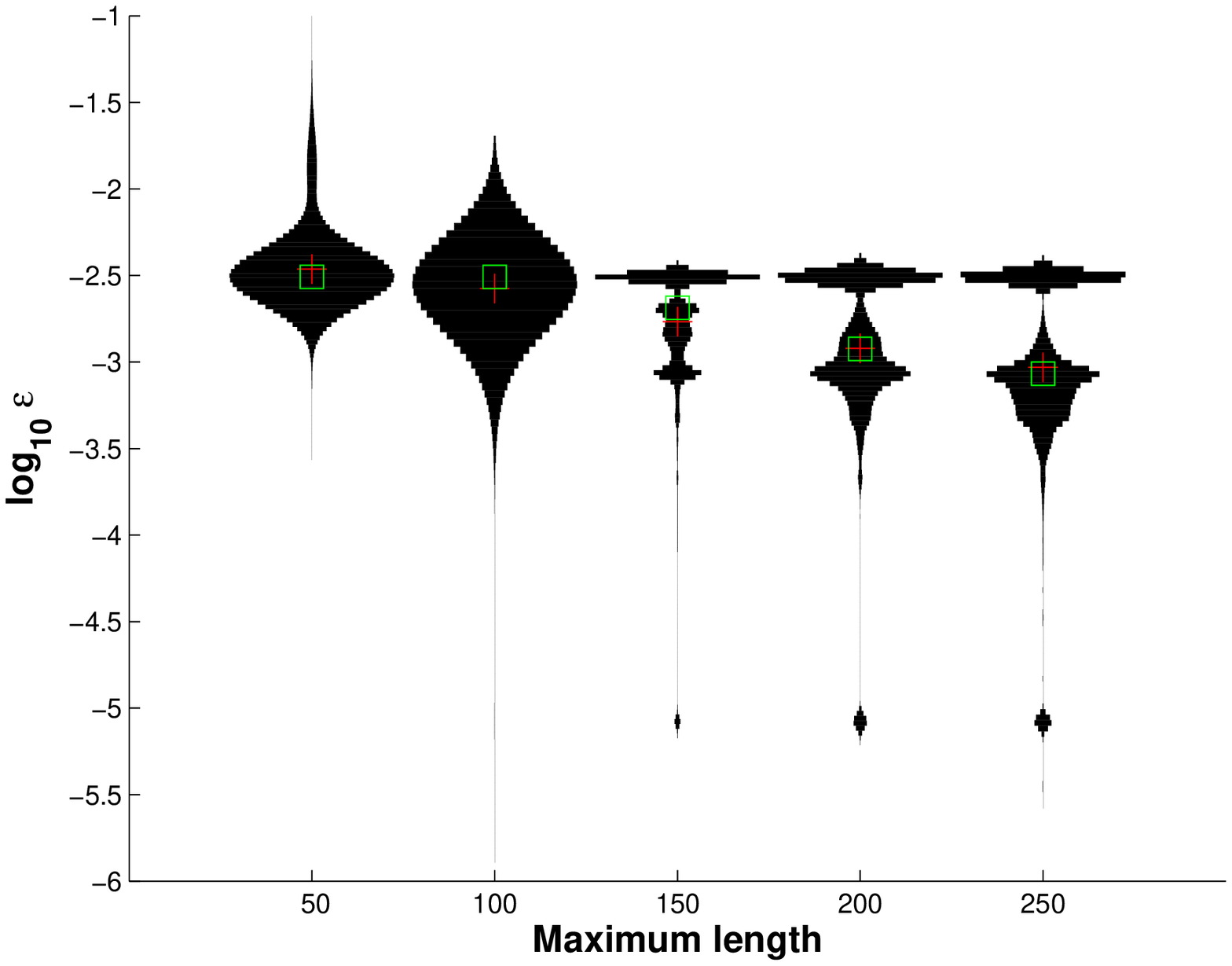}
\begin{pspicture}(0,-0.25)(12,0)
    \rput(1,-0.125){a)} \rput(6,-0.125){b)} \rput(13.0,-0.125){c)}
\end{pspicture}
     \caption{Violin plots showing the distribution of the best values found in all the executions for: a) All EDAs variants without local optimizer ($14400$ runs), b: All EDAs variants with local optimizer  ($14400$ runs), c: EDAs with local optimizer, recoding type II, and that use partial sampling type II ($300$ runs).}
    \label{fig:VIO}
    \end{center}
\end{figure*}

 We further investigate the behavior of the different EDA variants. Figure~\ref{fig:VIO} shows the violin plots \cite{Hintze_and_Nelson:1998} with the distribution of the best values found in all the executions for: a) All EDA variants without local optimizer ($14400$ runs), b) All EDA variants that incorporate the greedy search  ($14400$ runs), c) EDAs with local optimizer, recoding type II, and that use partial sampling type II ($300$ runs). Each violin plot shows a  histogram  smoothened using a kernel density with Normal kernel.  The mean and median are shown as red crosses and  green squares, respectively.

 In Figure~\ref{fig:VIO}, the modes  of the Normal distribution indicate the existence of a local optimum with a very wide basin of attraction. This local optimum has value $log_{10}f(\epsilon) = -2.50785$ and the majority of the EDA runs can be trapped in this value. Differences between the EDAs due to the application of the greedy method can be appreciated for $n=200$ and $n=250$ (Figures~\ref{fig:VIO}a) and~\ref{fig:VIO}b)). Also, Figure~\ref{fig:VIO}c) reveals how a particular combination of the EDA's parameters can improve the results of the search. This is shown in detail in Table~\ref{tab:F5RES} that comprises  all EDA variants that reached one of the best solutions in at least one of the $100$ runs.

\begin{table}[htb]
\scriptsize
\begin{center}
 {$ \begin{array}{|r|r|r|r|r||r|r|r|r|r|r|} \hline 
  L & tf& t \lambda & ts & pm &50 & 100 &150 & 200 &250 & tot \\ \hline\hline 
  0&3& 1& 0& 0& 0& 0& 1& 1& 0& 2\\ 
  0&3& 1& 1& 0& 1& 0& 1& 4& 0& 6\\ 
  0&3& 1& 1& 1& 0& 0& 1& 2& 0& 3\\ 
  0&3& 1& 1& 2& 0& 0& 1& 0& 0& 1\\ 
  0&3& 1& 2& 0& 0& 0& 0& 2& 0& 2\\ 
  0&3& 1& 2& 1& 0& 0& 0& 2& 0& 2\\ 
  0&3& 1& 2& 2& 0& 0& 0& 3& 1& 4\\ 
  0&3& 2& 0& 0& 0& 0& 0& 1& 0& 1\\ 
  0&3& 2& 1& 0& 0& 0& 4& 3& 0& 7\\ 
  0&3& 2& 1& 1& 0& 0& 0& 1& 0& 1\\ 
  0&3& 2& 1& 2& 0& 0& 1& 1& 0& 2\\ 
  0&3& 2& 2& 0& 0& 1& 0& 1& 0& 2\\ 
  0&3& 2& 2& 1& 0& 1& 1& 1& 0& 3\\ 
  0&3& 2& 2& 2& 0& 0& 2& 3& 0& 5\\ \hline
  1&2& 2& 1& 2& 0& 0& 1& 0& 0& 1\\ 
  1&3& 0& 1& 2& 0& 0& 1& 0& 0& 1\\ 
  1&3& 1& 1& 0& 0& 0& 0& 2& 0& 2\\ 
  1&3& 1& 1& 2& 0& 0& 0& 1& 0& 1\\ 
  1&3& 1& 2& 0& 0& 0& 9& 12& 0& 21\\ 
  1&3& 1& 2& 1& 0& 2& 14& 21& 0& 37\\ 
  1&3& 1& 2& 2& 0& 1& 18& 19& 0& 38\\ 
  1&3& 2& 1& 0& 0& 0& 0& 4& 0& 4\\ 
  1&3& 2& 1& 1& 0& 0& 1& 1& 0& 2\\ 
  1&3& 2& 1& 2& 0& 0& 1& 0& 0& 1\\ 
  1&3& 2& 2& 0& 0& 0& 4& 20& 0& 24\\ 
  1&3& 2& 2& 1& 0& 0& 2& 9& 0& 11\\ 
  1&3& 2& 2& 2& 0& 0& 4& 11& 1& 16 \\\hline
\end{array}$}
\caption{EDAs variants that obtained one of the best solutions at least one in the $100$ experiments. $L$: Local optimizer, tf: Type of function and representation, $t\lambda$: type of $\lambda$ value, $ts$: type of sampling, $pm$: probabilistic model.}
\label{tab:F5RES}
\end{center}
\end{table}

 There are a number of commonalities between the best EDA variants included in Table~\ref{tab:F5RES}. Except in one case, all EDAs use recoding of type II. Similarly, except in one case, in all the variants $\lambda \in \{ 0.01, 0.05\}$. Except in two cases, the sampling method selected was partial sampling. The application of the local optimizer notably improved the results for $n \in \{150, 200\}$ but in terms of the best solution found it did not have an important influence for the other values of $n$. 

 As a summary, we recommend to use an EDA that incorporates the greedy search, and uses partial sampling of type II and the 1-order Markov model since it is less complex than the tree and results achieved by the two models are similar.
  
\subsection{Improvement over other search methods}

 As a final validation of our method, we compare it with the results achieved using a random search, the greedy local optimizer, and the GA introduced in \cite{Mcdonald_and_Katzgraber:2013}. We compare our best EDA variant as described in the previous section. For the random search, we randomly selected $10000$ solutions and selected the best solution according to function $\bar{f}, \lambda = 0.01$. The same experiment was repeated $100$ times to select the $100$ ``best'' solutions for $n \in \{50,100,150,200,250\}$.

\begin{figure}[htb]
  \begin{center}
 \includegraphics[width=6.0cm]{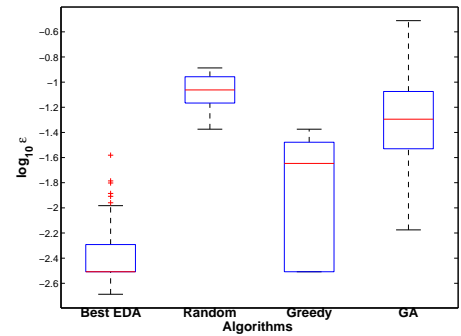}
  \caption{Comparison between the Best EDA variant, a random search, the greedy local search and the GA for $n=50$.}
    \label{fig:BPn50}   
 \end{center}
\end{figure}

\begin{figure}[htb]
  \begin{center}
 \includegraphics[width=6.0cm]{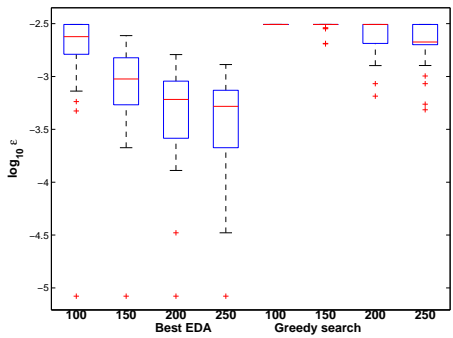}
  \caption{Comparison between the Best EDA variant and the greedy local search  $n \in \{100,150,200,250\}$.}
    \label{fig:ALLBP}   
 \end{center}
\end{figure}

 A similar procedure was followed for the greedy local search. The local optimizer was applied to each of the $10000$ solutions until no improvement was possible. For the GA, we used the results of $100$ GA runs used for the work published in \cite{Mcdonald_and_Katzgraber:2013}. Since these results were obtained using solutions of different length, and with a different number of evaluations, care must be taken to interpret the differences. We only compare the GA results with the other algorithms for $n=50$. Similarly, the results of the random search search were very poor for $n>50$ and we only include them in the comparison for $n=50$. Results are shown in Figure~\ref{fig:BPn50}. The results  of the comparison between the EDA and the greedy search for $n>50$ are shown in Figure~\ref{fig:ALLBP}.  In the boxplots, the central mark is the median, the edges of the box are the 25th and 75th percentiles, the whiskers extend to the most extreme data points not considered outliers, and outliers are plotted individually.

 It can be seen in Figures~\ref{fig:BPn50} and~\ref{fig:ALLBP} that the EDA significantly outperforms all the other methods. Furthermore, as $n$ increases the algorithm is able to scale and find better solutions. 
  
\section{Conclusions}  \label{sec:CONCLU}
  
  In this paper we have proposed the use of different EDA variants for the quasiparticle braid problem. We have shown that the fitness function and general evolutionary optimization approach initially introduced with GAs, can be successfully extended by the application of EDAs.  The best braids obtained with our EDAs have lengths  up to $9$ times shorter than those expected from braids of the same accuracy obtained with the Solovay-Kitaev algorithm and had not been previously reported to be found by the GA approach. We have also proposed three different methods to enhance the behavior of EDAs. Our results show that although the local optimizer improves the results of the EDA,  it is not able to reach solutions of similar quality when applied alone. Decoding, and particularly partial sampling, can be used as effective methods when dealing with other real-world problems as a way to improve usability of the representation and avoid disrupting complex solutions, respectively.

By means of analyzing the Boltzmann distribution we have shown that some of the problem characteristics are translated into statistical regularities of the Boltzmann distribution. In the future we plan to extend this analysis to try to extract more problem specific information that can be useful for designing more effective search methods. Other evolutionary algorithms that use models able to represent higher order dependencies, such as  Bayesian networks \cite{Larranhaga_and_Lozano:2002r}, Markov networks \cite{Shakya_and_Santana:2012}, and factor graphs \cite{Helmi_et_al:2014}, could be applied.  We also plan to address other braid problems of higher difficulty.

\section{Acknowledgments}

R. Santana has been partially supported by the Saiotek and Research Groups 2013-2018 (IT-609-13) programs (Basque Government), TIN2013-41272P (Ministry of Science and Technology of Spain), COMBIOMED network in computational bio-medicine (Carlos III Health Institute), and by the NICaiA Project PIRSES-GA-2009-247619 (European Commission).  H. G. Katzgraber acknowledges support from the NSF (Grant No. DMR-1151387).

\bibliographystyle{abbrv}
\bibliography{Thesbib}  

\end{document}